\documentclass[reprint,prl,aps,showpacs,footinbib,twocolumn]{revtex4-1}

\usepackage{amssymb}
\usepackage{graphicx}
\usepackage{amsmath}
\usepackage{times}
\usepackage{color}
\usepackage{setspace}
\usepackage{bm}
\usepackage{float}
\usepackage{subfigure}
\usepackage{hyperref}
\usepackage{braket}

\usepackage{mathtools}
\usepackage{newfloat}
\DeclareFloatingEnvironment[name={Figure S}]{suppfigure}

\definecolor{purple}{rgb}{0.6,0.1,1}
\allowdisplaybreaks
\begin{document}

\title{Suppressing the loss of ultracold molecules via the continuous quantum Zeno effect}

\author{B. Zhu$^{1}$, B. Gadway$^{1}$, M. Foss-Feig $^{2}$, J. Schachenmayer$^{1}$, M. L. Wall $^{1}$, K.~R.~A. Hazzard$^{1}$,  B. Yan$^{1}$, S.~A. Moses$^{1}$, J.~P. Covey$^{1}$,  D.~S. Jin$^{1}$, J. Ye$^{1}$, M. Holland$^{1}$, A.~M. Rey$^{1}$ }\email{arey@jilau1.colorado.edu}
\affiliation{$^{1}$JILA, NIST, Department of Physics, University of Colorado, 440 UCB, Boulder, CO 80309, USA}
 \affiliation{$^{2}$ JQI, NIST,  Department of Physics, University of Maryland, College Park, MD 20742, USA}
\date{\today}

\begin{abstract}
  We investigate theoretically the suppression of two-body losses when
  the on-site loss rate is larger than all other energy scales in a
  lattice. This work quantitatively explains the
  recently observed suppression of chemical reactions between two
  rotational states of fermionic KRb molecules confined in
  one-dimensional tubes with a weak lattice along the tubes [Yan {\it
    et al.}, Nature \textbf{501}, 521-525 (2013)].  
    New loss rate measurements performed for different lattice parameters but under controlled initial conditions allow us to show that the loss suppression is a consequence of the combined effects of
  lattice confinement and the continuous quantum Zeno effect. A key
  finding, relevant for generic strongly reactive systems, is
  that while a single-band theory can qualitatively describe the data,
  a quantitative analysis must include multiband effects. 
  Accounting for these effects reduces the inferred molecule filling fraction by a factor of five.  
A rate equation can describe much of the data, 
 but to properly reproduce the loss dynamics with a fixed filling fraction for all lattice parameters 
  we develop a mean-field model and  benchmark it with numerically exact
  time-dependent density matrix renormalization group
  calculations. 

 \end{abstract}
\pacs{03.65.Xp,67.85.-d,37.10.Jk,37.10.Pq}

 \maketitle

\begin{figure}[!htbp]
\begin{center}
 \includegraphics[width=0.47\textwidth]{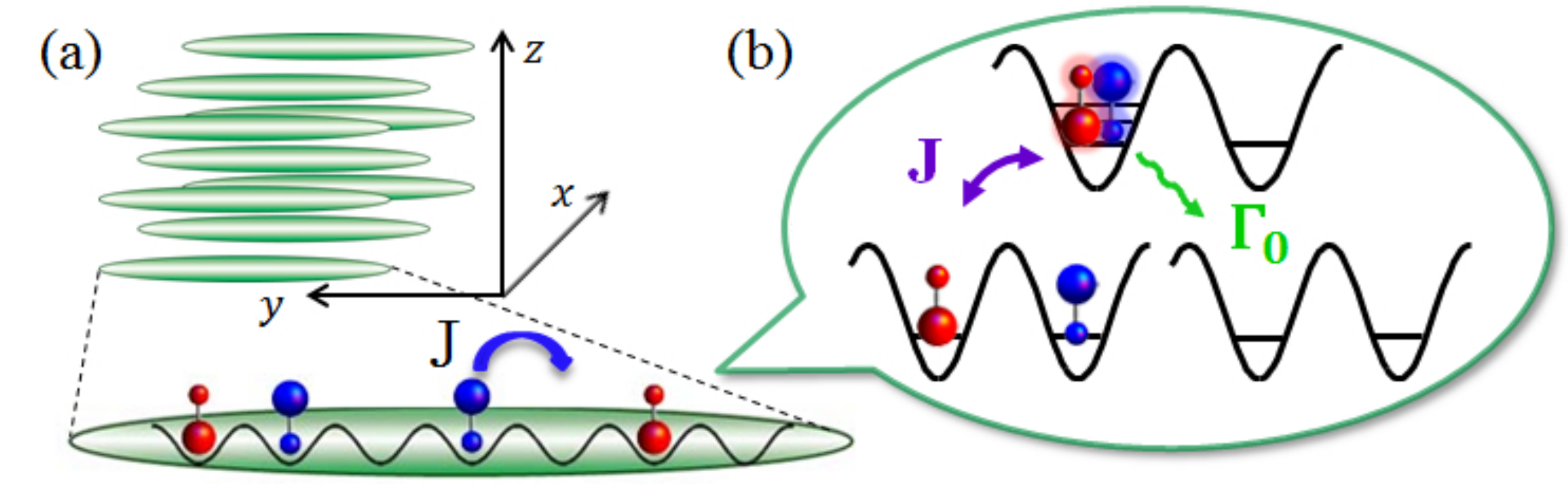}
\end{center}
  \caption{(a) A 50:50 mixture of fermionic KRb molecules in two rotational states,  $\ket{0,0}$ (red) and  $\ket{1,-1}$ (blue),  is prepared in a deep 3D lattice, which is suddenly made shallow along one dimension ($y$).  Along $y$, molecules tunnel with a rate $J/\hbar$ and have a large on-site loss rate $\Gamma_0$ because of chemical reactions. (b) In the Zeno regime,  $\hbar\Gamma_0\gg J $, doubly occupied sites are only virtually populated, and the loss occurs at a significantly slower rate $\Gamma_{\rm eff}\ll \Gamma_0$ for  molecules on adjacent
  sites. For KRb, a multiband analysis of this process is required for all experimental lattice parameters.}\label{fig:zeno}
\end{figure}

Ultracold molecules have tremendous applications ranging from  quantum many-body physics~\cite{Trefzger,baranov,lahayephysics} and quantum information processing~\cite{DeMille} to precision measurements~\cite{Zelevinsky} and  ultracold chemistry~\cite{carr:cold_2009}.  However, fast inelastic two-body losses -- as occur for KRb with exothermic chemical reactions -- can limit molecule lifetimes~\cite{silke2010,ni2010,miranda2011}, and have been considered a fundamental limitation. Recent experiments with KRb molecules~\cite{Bo2013} have  reported  an inhibition of losses when the molecules are confined in an array of one-dimensional tubes with a superimposed  axial optical lattice along the tubes  (Fig.~\ref{fig:zeno}). Similar loss suppression by strong dissipation was previously observed in bosonic Feshbach molecules~\cite{rempe2008}. Extending  the molecules' lifetime over timescales much longer than those determined by tunneling opens a path for the exploration of itinerant magnetism and other  many-body phenomena arising from the interplay between dipolar interactions and motion, even in these  in highly reactive systems.

 Free KRb molecules  react rapidly.  In a lattice, the  two-body inelastic collision rates are larger than all other lattice energy scales, including the band separation energy. 
 Consequently, this system is an example of a strongly correlated system that defies simple treatment in terms of single-particle physics. As such, description of the loss suppression based on the assumption that inelastic interactions do not affect the single-particle wave-functions is incorrect~\cite{mark2011}. Evidence of this issue was reported in Ref.~\cite{Bo2013} where a heuristic ``single-band" treatment of the losses  was found to significantly  overestimate the molecule filling fraction $f$. 

\begin{figure*}[!htbp]
  \begin{center}
  \includegraphics[width=0.92\textwidth]{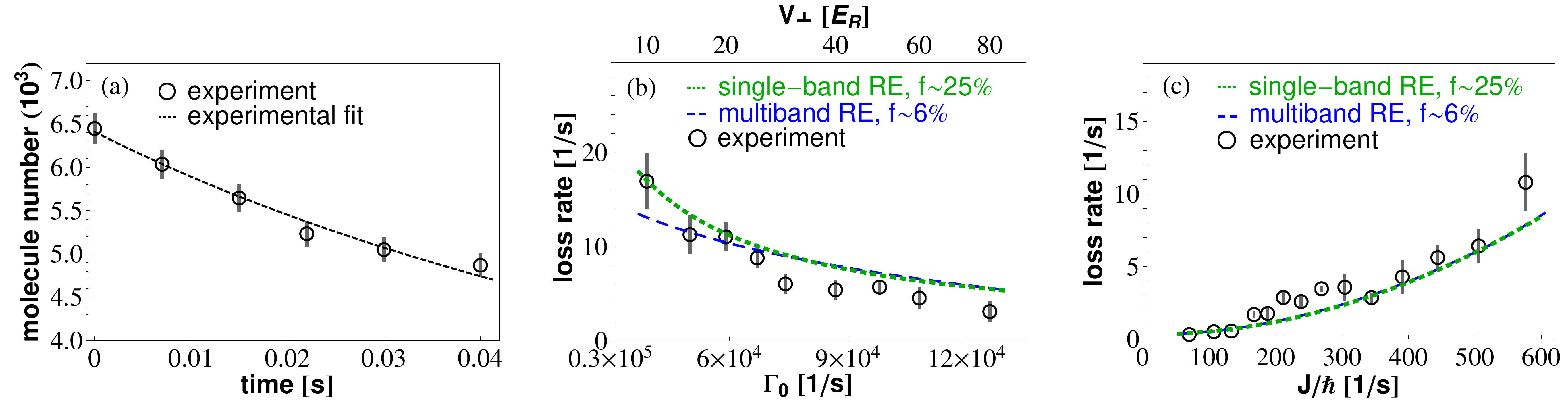}
 \end{center}
\caption{ (a) Measured number loss of $\ket{\downarrow}$ molecules for an axial (transverse) lattice depth of $V_y=5\,E_R$ ($V_\perp=25\,E_R$) (circles) and best fit using a rate equation (RE), Eq.~\eqref{eqn:numberloss} (black dashed line). 
(b) Number loss rate, $\kappa$, as a function of $\Gamma_0$  (fixing $J\approx570~$Hz and varying the bare on-site rate via $V_\perp$).  (c) Number loss rate, $\kappa$, versus $J$ for fixed $\Gamma_0\approx87$~kHz (varying $V_y$ and adjusting $V_{\perp}$  accordingly).  $V_y$ ($V_{\perp}$) was varied from 5 to $16\,E_R$ ($20$ to $40\,E_R$). Black circles are experimental measurements (error bars represent one standard error).  Green short-dashed lines show solutions of the RE Eq.~\eqref{eqn:numberlosssb} using an effective loss rate $\Gamma_{\rm{eff}}$ (single-band approximation). The blue long-dashed line shows the multiband RE  using $\tilde{\Gamma}_{\rm{eff}}$ in Eq.~\eqref{eqn:numberlosssb}. The  multiband  and single-band RE results  were obtained by fixing the filling fraction to be $ 6\%$, and $25\%$ respectively. Panels (b) and (c) directly manifest the continuous quantum Zeno effect: in (b) the measured loss rate $\kappa$ decreases with increasing on-site $\Gamma_0$; in (c) a fit to the experimental data  supports $\kappa\propto J^2$, with a $\chi^2$ (sum of the squared fitting errors) several times smaller than for a linear fit.}  
\label{fig:zenoex}
\end{figure*}

 In this Letter, we develop a theoretical description of the  dissipative dynamics  that 
 non-perturbatively includes three dimensional multi-band effects. Our analysis  allows  us to
  attribute the observed loss suppression  to  the continuous quantum Zeno effect~\cite{misra:zeno_1988,itano:quantum_1990,PhysRevLett.87.040402,Han2009,Ripoll2009} -- a suppression of coherent transitions  due to strong dissipation -- and to   generalize previous single-band treatments~\cite{Ripoll2009,baur:two-body_2010} to the  strongly dissipative regime.   We perform systematic measurements of  the KRb lifetime under controlled and reproducible lattice conditions that allow us to validate the calculations. The observed  dependence  of the loss rate on lattice parameters is consistent with  Ref.~\cite{Bo2013} and is fully reproduced by the multiband theory. Moreover, the inclusion of multiple bands reduces the determined filling $f$ by a factor of ${\sim} 5$, giving results consistent with the filling predicted by Ramsey spectroscopy measurements of molecules pinned in a 3D lattice and prepared under similar initial conditions~\cite{Bo2013,Hazzard2013}. 

The multiband calculations are first applied  to derive a simple rate equation (RE) for  two-body losses, which assumes instantaneous redistribution of molecules between collision events. We show that the RE  can describe  the measured dissipative  dynamics fairly well  over  a broad range of lattice parameters, but fails for the deepest lattice configurations. 
 We develop a simple and unified theory capable of describing the loss dynamics in \emph{all} parameter regimes based on a mean-field (MF) approximation of the many-body master equation.  We  validate the MF formulation by comparing it to a numerically exact time-dependent density matrix renormalization group method (t-DMRG)~\cite{Vidal2004,Daley2004,White2004}, which we combine with a quantum trajectory technique~\cite{carmichael_open_1991,molmer_monte_1993,dum_monte_1992}.  The MF, t-DMRG, and experimentally  observed loss dynamics quantitatively agree.

{\it Experiment--}  The experiment begins  by loading ${\sim}10^4$  fermionic KRb ro-vibrational ground-state molecules, $|N=0,m_N=0\rangle$, into the lowest  band of a deep 3D cubic optical  lattice with lattice constant $a= 532$~nm.  Here, $N$ is the principal rotational quantum number and $m_N$ is the projection onto the quantization axis, which in our case is determined by an external magnetic field angled $45^{\circ}$ between the $x$ and $y$ lattice directions. 
We next  apply a $\pi/2$ microwave pulse to rotationally excite half of the molecules to $|N=1,m_N=-1\rangle$.  We  consider  $|0,0\rangle$ and $|1,-1\rangle$ as $\ket{\downarrow}$ and $\ket{\uparrow}$ components of a pseudo-spin $1/2$ system. We choose the lattice polarizations so that the tensor AC polarizabilities of $|0,0\rangle$ and $|1,-1\rangle$ are similar~\cite{brian}. 
However, a residual differential AC Stark shift introduces single-particle dephasing that results in a spin-coherence time for the entire sample of ${\sim}1$~ms. This  dephasing  allows us  to prepare  an incoherent  50:50 spin mixture of $\ket{\downarrow}$ and $\ket{\uparrow}$ by  holding the molecules in the deep lattice for 50~ms. Losses are then initiated by quickly ramping down the lattice depth in the $y$ direction  (within 1~ms) to allow tunneling. We measure the  number of
remaining molecules $\ket{\downarrow}$, i.e., $N_{\downarrow}(t)$, as  a function of the subsequent holding time in the lattice.

We experimentally determine the initial loss rate $\kappa$ by fitting $N_{\downarrow}(t)$ to the solution of a two-body loss RE of the form \begin{align}
\frac{dN_{\downarrow}}{dt}=-\frac{\kappa}{N_{\downarrow}(0)} [N_{\downarrow}(t)]^2 , \label{eqn:numberloss}
\end{align}  with $N_\downarrow(0)$ the initial number of $\ket{\downarrow}$ molecules. A typical experimental fit   is shown in Fig.~\ref{fig:zenoex}(a).  
To avoid  the  saturation of the losses that originates from the finite number of molecules per tube ($\sim$6 per tube on average),  which cannot be captured by the RE, we fit only up to times when $\sim25\%$ of the molecules are lost (see Supplementary Material).

The loss rate $\kappa$  in general depends on the tunneling rate $J/\hbar$ and the on-site ``bare'' loss rate $\Gamma_0$. If a single-band approximation is used,  the on-site bare loss rate $\Gamma_0$   is given by~\cite{Ripoll2009}
\begin{align}
\label{inter} \textstyle\Gamma_0=\beta^{(3D)}\int|W({\bf x})|^4  \mathrm{d}^3 {\bf x},
\end{align} where $W({\bf x})$ is the lowest-band single-particle 3D Wannier orbital. The two-body loss rate coefficient for molecules in  $|0,0\rangle$ and $|1,-1\rangle$, $\beta^{(3D)}$, was  measured to be $\beta^{(3D)}=9.0(4)\times 10^{-10}$~cm$^{3}$~s$^{-1}$~\cite{Bo2013}.

To experimentally extract the  dependence of  $\kappa$  on $\Gamma_0$ and the single-particle hopping energy $J$,  we perform  similar measurements to those reported in Ref.~\cite{Bo2013}. However, here we ensure reproducibility of the initial conditions to fix $f$ for all lattice conditions. To measure the $\Gamma_0$-dependence  we set $V_y=5\,E_R$, which fixes $J$, and then tune $\Gamma_0$ by modifying  $V_\perp$ [Fig.~\ref{fig:zenoex}(b)].  Here, $E_R=\hbar^2\pi^2/2ma^2$ is the recoil energy and $m$ is the KRb mass. To study the $J$-dependence, we vary $V_y$ while simultaneously adjusting $V_\perp$ to keep $\Gamma_0$ fixed [Fig.~\ref{fig:zenoex}(c)]. The loss rate $\kappa$ is found to depend quadratically on $J$  for fixed $\Gamma_0$ and to decrease  with increasing  $\Gamma_0$ for fixed $J$.  
This scaling is consistent with the continuous quantum Zeno effect, as we now explain.

{\it Single-band rate equation--}A simple way to understand the loss suppression is to consider two opposite spin particles in a double well, $\ket{\uparrow,\downarrow}$.
Left and right sides in this notation represent left and right wells  [Fig.~\ref{fig:zeno}(b)]. When two molecules occupy the same site the  singlet component   decays with rate $\Gamma_0$, while the decay of  the triplet component
is suppressed by the centrifugal barrier in odd partial-wave  channels~\cite{miranda2011}. Consequently, the loss rate is determined by  $J_s=\sqrt{2} J$, which is the tunneling computed after projecting the initial wavefunction into the singlet state $\ket{s}=(\ket{\uparrow,\downarrow}-\ket{\downarrow,\uparrow})/\sqrt{2}$.

When $\hbar \Gamma_0\gg J_s$,   second-order perturbation theory can be applied and gives
a net $\ket{\downarrow}$ loss rate of  $4 \Gamma_{\text{eff}}$ with  $\Gamma_{\text{eff}}=\frac{2(J/\hbar)^2}{\Gamma_0}$.
This  loss rate can be connected to number loss dynamics with a RE,
$\frac{dn_{j\downarrow}}{dt}=-4q \Gamma_{\text{eff}} n_{j+1\uparrow} n_{j\downarrow}$,
 where $n_{j\downarrow}$ is the number of $\ket{\downarrow}$ molecules at site $j$ and $q$ is the number of nearest neighbor lattice sites ($q=2$ for tunneling  along the tube direction)~\cite{baur:two-body_2010}.  
 Assuming a uniform distribution, the 50:50 mixture implies $n_{j+1\uparrow}=n_{j\downarrow}=n_{\downarrow}$ and
\begin{align}
\frac{dn_{\downarrow}}{dt}=-8\Gamma_{\text{eff}} [n_{\downarrow}(t)]^2\quad {\rm{or}}\quad
\frac{dN_{\downarrow}}{dt}=-\frac{\kappa_{SB}}{N_{\downarrow}(0)} [N_{\downarrow}(t)]^2 , \label{eqn:numberlosssb}
\end{align}
where $\kappa_{SB}=8\Gamma_{\text{eff}}n_{\downarrow}(0)$. All parameters are known except the filling fraction   $f=2n_{\downarrow}(0)$. The RE  assumes that the loss rate depends only on the \emph{average} density.  This assumption is valid when the redistribution of density after a loss process occurs faster than the typical time between losses ($J\gg\Gamma_{\text{eff}}$)~\cite{fnote1}.

This simplified single-band model  qualitatively  reproduces the measured dependence of $\kappa$ on lattice parameters [green-short-dashed line in Figs.~\ref{fig:zenoex}(b)-(c)]. However,  since $\Gamma_0$ is  larger than the band gap (e.g., 4 times larger for a $V_y=5~E_R$ and $V_\perp=40~E_R$ lattice), this single-band theory is known to be inadequate. Moreover,
in order to fit the experiment, the single-band theory requires $f\approx 25\%$, which is known to be  inconsistent with estimates of the filling  $f\lesssim 10 \%$ from Ramsey spectroscopy procedures \cite{Hazzard2013,chotia:long-lived_2012}. Resolution of this discrepancy requires including  multiple single-particle bands, which are admixed by strong two-body losses.

{\it Multiband rate equation--} As shown in the Supplementary Material, a single-band model overestimates $\Gamma_0$, predicting it to be  larger than the band gap. Incorporating higher bands decreases $\Gamma_0$ and  hence decreases the  $f$ estimated from experiment (since the effective loss rate is inversely proportional to $\Gamma_0$).
We extract a renormalized effective loss rate by numerically computing the loss of two molecules trapped in a double well along $y$. We expand the non-Hermitian Hamiltonian $\hat{H}=\hat{H}_{0}-i\hbar\beta^{(\mathrm{3D})}\delta_{\mathrm{reg}}\left(\mathbf{r}\right)/2$, where $\delta_{\mathrm{reg}}\left(\mathbf{r}\right)=\delta\left(\mathbf{r}\right)\left(\partial/\partial r\right)r$ is a regularized pseudopotential~\cite{huang1957} and $\hat{H}_{0}$ the single-particle Hamiltonian, in the 3D Wannier function basis.  This model accounts for interaction-mediated  band excitations in all three dimensions.  We initialize the system with two molecules in the singlet $|s\rangle$ and infer the effective loss rate by fitting the norm decay to $\exp(-4\tilde\Gamma_{\mathrm{eff}}t)$.  Convergence is achieved with 6 bands in each dimension.


Surprisingly, as shown in Figs.~\ref{fig:zenoex}(b--c), both effective loss rates $\Gamma_{\text{eff}}$ and ${\tilde \Gamma}_{\text{eff}}$ scale similarly with $\Gamma_0$ and $J$.   This similarity explains why qualitative experimental signatures of Zeno suppression expected from a single-band model survive even though such a model is invalid.  However, the multiband ${\tilde \Gamma}_{\text{eff}}$ is $\sim$5 times larger than $\Gamma_{\text{eff}}$.  Once these effective loss rates are calculated, the only free parameter to fit the experimental measurements is the filling $f$, which was fixed to be the same for all data shown in Figs.~\ref{fig:zenoex}(b--c).  The $\sim5$ times faster loss rate from the multiband model leads to a $\sim 5$ times smaller filling fraction of $f=6\%$ [Figs.~\ref{fig:zenoex}(b--c), blue-long-dashed line] compared to the grossly overestimated $25\%$ extracted using $\Gamma_{\text{eff}}$ [Figs.~\ref{fig:zenoex}(b--c) green-short-dashed lines].  The inadequacy of the single-band model to extract the correct filling fraction, and the success of the multiband model, are key results of this work.


 {\it Mean-field and DMRG --}The RE, with parameters extracted from the multiband model, describes the experimental observations fairly well at intermediate $V_\perp$, but deviates from them for the largest $V_\perp$.  We attribute these deviations
   to the suppression of tunneling at the cloud's edges due to the energy mismatch between adjacent sites in the harmonic potential generated by the  lattice beams. By inhibiting transport, this effect invalidates the assumption that molecules are redistributed rapidly between loss events, and therefore the losses are  not determined exclusively
by the average density but depend on the detailed dynamical redistribution of molecules.

Although this redistribution is absent from the RE, it can be accounted for by  solving a master equation with a density matrix,   $\hat \rho$, projected into the states with at most  one molecule per site after adiabatic elimination of doubly occupied states. We  keep terms up to order $\Gamma_{\text{eff}}$~\cite{Ripoll2009}, and we  simultaneously account for multiband effects by replacing the single-band  $\Gamma_{\rm{eff}}$ by the renormalized loss rate extracted from the multiband double well solution, obtaining
\begin{align}
\label{master}  \frac{d}{dt} {\hat \rho} &= -\frac{{\rm i}}{\hbar}[ {\hat H_0} , {\hat \rho} ] +  {\mathcal L}{\hat\rho}\, .
\end{align}
Here
${\hat H_0}=-J\sum_{j,\sigma}  (\hat{c}^\dagger_{j\sigma }\hat{c}_{j+1\sigma} + h.c)+ \sum_{j,\sigma} V_j^\sigma \hat{c}^\dagger_{j\sigma }\hat{c}_{j\sigma }$,
${\mathcal L}{\hat\rho}= \frac{1}{2} \sum_{j}  \left[2\hat{A}_{j} {\hat \rho} \hat{A}_{j}^\dagger -
{\hat \rho} \hat{A}_{j}^\dagger  \hat{A}_{j}-\hat{A}_{j}^\dagger  \hat{A}_{j} {\hat \rho} \right]$~\cite{fnote}, 
and
 $V_j^\sigma=\frac{1}{2} m\omega_\sigma^2 j^2 a^2 $ is the parabolic trapping potential felt by molecules in state $\sigma$ at site $j$. The average trap frequency $(\omega_\uparrow+\omega_\downarrow)/2$ varies between $\approx 2\pi \times (15 -40) \:\rm{Hz}$ for  the experimental range of $V_\perp$. The $\sigma$-dependence is due to residual
differential AC Stark shifts between the two rotational states. 
${\mathcal L}$ is a Lindblad superoperator that accounts for losses, and the jump operators are
$ \hat{A}_j=\sqrt{2\tilde\Gamma_{\text{eff}}}\Big[(\hat{c}_{j\uparrow} \hat{c}_{j+1\downarrow}+\hat{c}_{j\uparrow}\hat{c}_{j-1\downarrow})-(\hat{c}_{j\downarrow}\hat{c}_{j+1\uparrow}+\hat{c}_{j\downarrow}\hat{c}_{j-1\uparrow})\Big].
$
 We have checked the validity of the renormalized single-band model by confirming that  it reproduces the dynamics  of the multiband problem for the case of  two molecules in four wells.

  \begin{figure}
  \begin{center}
    \includegraphics[width=0.49\textwidth]{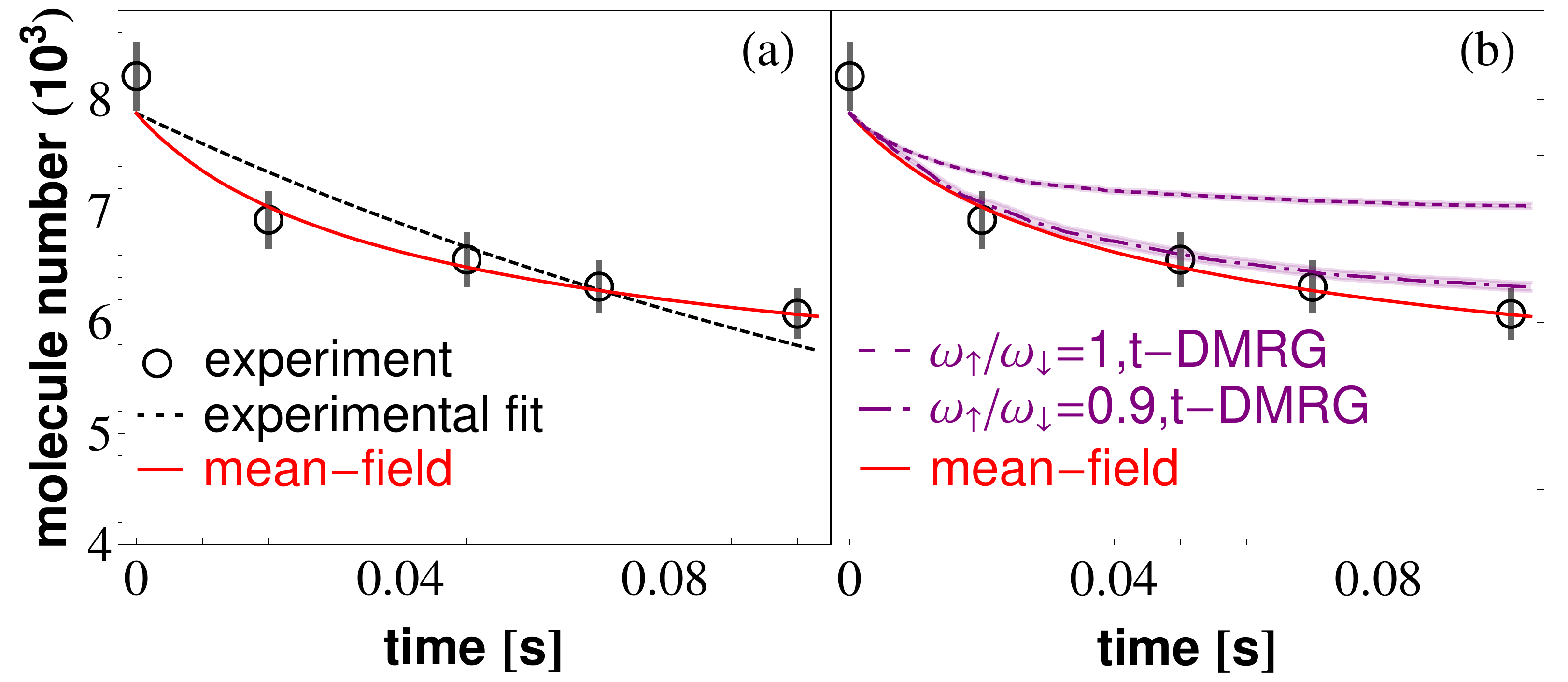}
 \end{center}
  \caption{ Comparison of experimental loss dynamics for the deepest considered lattice to MF and t-DMRG calculations. (a) Molecule loss vs.~time for $V_\bot=80 \, E_R$ and $V_y=5 \, E_R$ [Identical  conventions/conditions to Fig.~\ref{fig:zenoex}(b)]. The MF matches the experimental data better than the RE (experimental fit). (b) Comparison of t-DMRG simulations ($\chi_{\rm{MPS}}=128$, $2000$ trajectories) to MF, for two different cases: i) an identical trap for the two spin states with trap frequency $\omega_\downarrow=\omega_\uparrow=2\pi\times 38 \:{\rm Hz}$; and ii) slightly different trap frequencies $\omega_\downarrow=2\pi \times 38 \:{\rm Hz}, \omega_\uparrow=2\pi\times 34.2\: {\rm Hz}$. Shaded areas indicate the standard error of the mean.}
\label{compara}
\end{figure}

To solve Eq.~\eqref{master} we map the hardcore fermions onto hardcore spin-1/2  bosons~\cite{Sigrist1997}, and then use a mean-field ansatz $\hat{\rho}=\prod_j \tilde{\rho}_{j}$ with $\tilde{\rho}_{j} \equiv \sum_{\alpha,\beta=\{\uparrow,\downarrow,0\}} {\rho}_j^{\alpha,\beta} |\alpha\rangle \langle \beta|$.
Here, $\tilde{\rho}_{j}$ is the reduced projected density matrix at site $j$, and $\uparrow,\downarrow,0$ label the three possible local states of spin up, down, and the vacuum, respectively. This ansatz leads to closed equations of motion for ${\rho}_j^{\alpha,\beta}$ [see Supplementary Material]. Due to the rapid dephasing of spin coherence resulting from $\omega_{\uparrow}\neq\omega_{\downarrow}$, we  set  ${\rho}_j^{\sigma,\sigma'\neq\sigma}=0$, which simplifies the equations further.
Although the MF treatment predicts no coherent tunneling for a pure Fock state, we initiate it by assuming non-zero particle/hole coherence $|{\rho}_j^{\sigma,0}|=1/2$. 

Fig.~\ref{compara}(a) shows the dynamics for the largest $V_\perp$, where the coherent tunneling is strongly suppressed by the large parabolic potential $\omega_{\uparrow/\downarrow}$.  
We see that the dynamics is poorly described by the RE, and the MF solution better describes the data.  Admittedly, the MF assumption is an extreme approximation precluding entanglement between parts of the system. In order to test its validity, we also solve  Eq.~\eqref{master} numerically  by combining t-DMRG algorithms~\cite{Vidal2004,Daley2004,White2004} with a stochastic sampling over quantum trajectories~\cite{carmichael_open_1991,molmer_monte_1993,dum_monte_1992,s_sp_2013}. The results, shown in Fig.~\ref{compara}(b),  are converged in the matrix product state dimension $\chi_{\rm{MPS}}$ and are therefore numerically exact.  The  differential stark  shift for the lattice parameters of Fig.~\ref{compara} ($\omega_{\uparrow}/\omega_{\downarrow}\sim0.9$)   gives rise to an effective spatially dependent magnetic field that disrupts  spin correlations generated during the dynamics.  
 In this case the data, t-DMRG, and MF (which explicitly ignores spin-correlations) agree up to the times used to extract loss rates from the data, when the contrast has decayed by $\sim 20\%$.  However, in the absence of a differential Stark shift ($\omega_{\downarrow}=\omega_{\uparrow}$), we note that the density calculated from t-DMRG saturates at a higher value than predicted by the MF theory, which we attribute to the growth of spin correlations in the absence of dephasing~\cite{FossFeig2012}. 


\begin{figure}
  \begin{center}
  \includegraphics[width=0.48\textwidth]{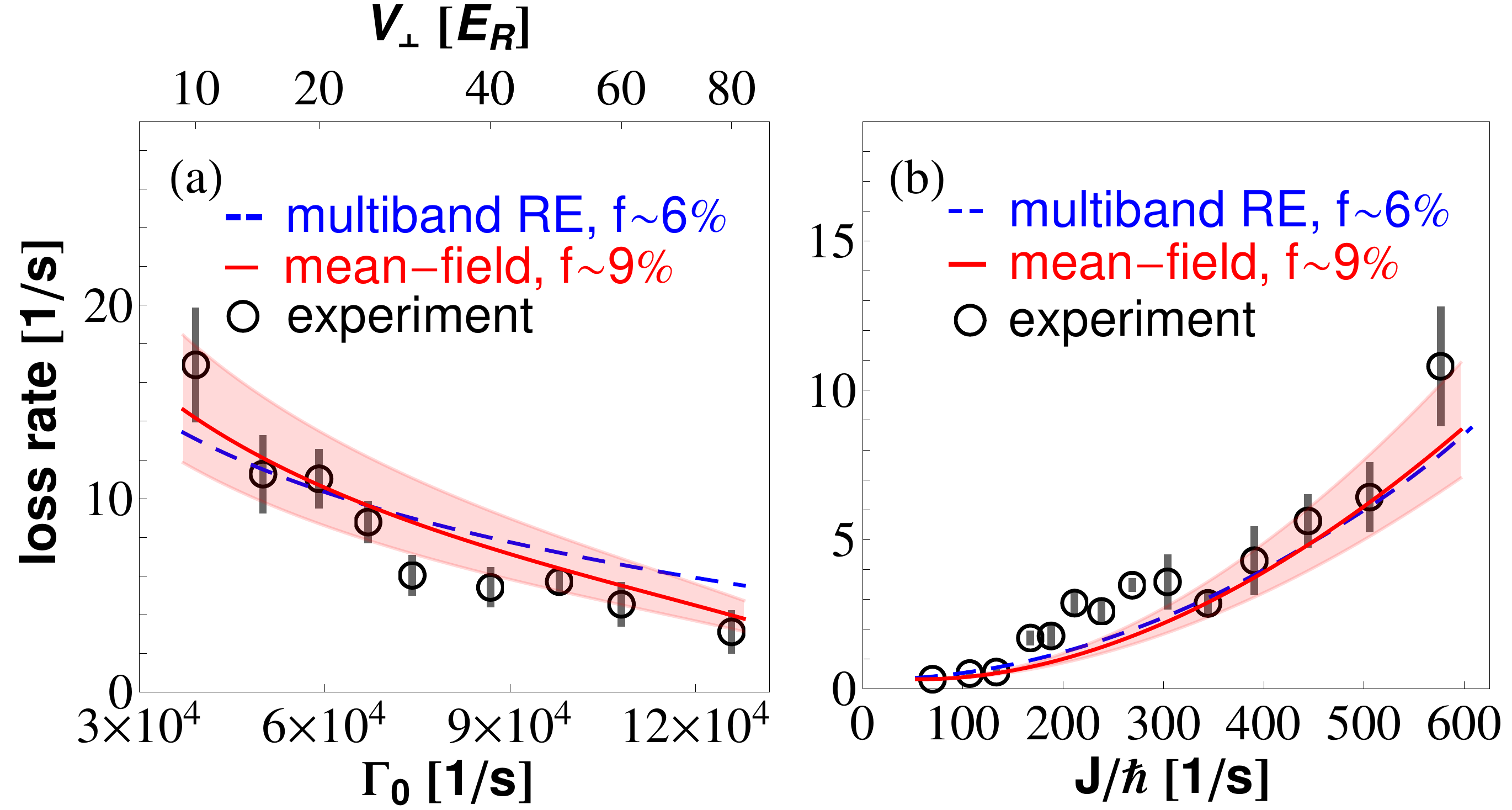}
 \end{center}
\caption{
(a) Number loss rate, $\kappa$, as a function of $\Gamma_0$  (same data as Fig. 2(b)).  (b) Number loss rate, $\kappa$, versus $J$ for fixed $\Gamma_0$ (same data as Fig. 2(c)). Black circles are experimental measurements.  Blue long-dashed and solid red lines show RE and MF solutions, respectively, using $\tilde{\Gamma}_{\rm{eff}}$ (multiband model). The shaded area  accounts for $\pm 2\%$ variations around the MF estimate of  $f \sim 9 \%$ arising from the uncertainty in the initial molecule distribution.}\label{fig:zenoex2}
\end{figure}

 {\it Mean-field vs. experiment--} With the validity of the MF established,  we  use it to model the experiment. 
 In the MF calculation, we assume that molecules are initially uniformly distributed within a shell with inner (outer) radius of 20 (50) lattice sites.  The shell distribution is expected because molecules are created from a Mott insulator of Rb and a band insulator of K.  Assuming only sites with one Rb and one K can yield molecules during STIRAP~\cite{silke2010,chotia:long-lived_2012}, sites in the trap center initially doubly occupied by Rb atoms are lost~\cite{silke2010}.  We then average over random initial configurations, since the experiments measure an ensemble of 1D tubes. Figures~\ref{fig:zenoex2}(a-b)   show the MF results (red line), where we used $f=9\%$ to match the experiment. This is slightly larger than that from the multiband RE (dashed blue line in Figures~\ref{fig:zenoex2}),  $f=6\%$, since the RE overestimates the loss rate by assuming instantaneous redistribution of the molecules. 
 
Since the molecule distribution in the experiment is known only approximately, we vary the shell width and find that the estimated MF filling fraction that best fits the experimental loss has a range  $f \sim 9 \pm 2\%$. The MF accounts better for  the dependence of the loss rate on $V_\perp$. Remaining deviations between MF and experiment are seen only for the shallowest $V_\perp$, where the transverse tunneling rate is only three times smaller than the axial one, which  may indicate the breakdown of 1D dynamics.

\textit{Conclusions}---The understanding of  the underlying physical mechanism responsible for the loss suppression in KRb 
 opens the path for laboratory  explorations of iconic models of quantum magnetism
combining motional and spin degrees of freedom, previously believed to be inaccessible due to losses. These include the extended $t$-$J$ model, predicted  to exhibit itinerant ferromagnetism,  $d$-wave superfluidity~\cite{Gorshkov2010, Kuns2011}, and topological phases~\cite{lahayephysics,Lewenstein}.  Our findings also extend to other dissipative  systems, such as alkaline earth atoms~\cite{Ludlow2011,Bishof2011b,Traverso2009,Lisdat2009} and other chemically reactive  molecular species~\cite{PhysRevA.81.060703}.

{ \it Acknowledgements:}
The authors thank A. Daley for useful discussions and   acknowledge funding from NIST, JILA-NSF-PFC-1125844, NSF-PIF, ARO, ARO-DARPA-OLE, and AFOSR. K.R.A.H., B.G. and M.F.-F. thank the NRC postdoctoral fellowship program for support. S.A.M. and J.P.C. acknowledge funding from NDSEG. A.M.R. and K.R.A.H. thank the KITP for its hospitality. This work utilized the Janus supercomputer, supported by NSF, NCAR and CU.
\bibliographystyle{apsrev}
\bibliography{bibdipole11b}

\begin{thebibliography}{44}
\expandafter\ifx\csname natexlab\endcsname\relax\def\natexlab#1{#1}\fi
\expandafter\ifx\csname bibnamefont\endcsname\relax
  \def\bibnamefont#1{#1}\fi
\expandafter\ifx\csname bibfnamefont\endcsname\relax
  \def\bibfnamefont#1{#1}\fi
\expandafter\ifx\csname citenamefont\endcsname\relax
  \def\citenamefont#1{#1}\fi
\expandafter\ifx\csname url\endcsname\relax
  \def\url#1{\texttt{#1}}\fi
\expandafter\ifx\csname urlprefix\endcsname\relax\def\urlprefix{URL }\fi
\providecommand{\bibinfo}[2]{#2}
\providecommand{\eprint}[2][]{\url{#2}}

\bibitem[{\citenamefont{Trefzger et~al.}(2011)\citenamefont{Trefzger, Menotti,
  Capogrosso-Sansone, and Lewenstein}}]{Trefzger}
\bibinfo{author}{\bibfnamefont{C.}~\bibnamefont{Trefzger}},
  \bibinfo{author}{\bibfnamefont{C.}~\bibnamefont{Menotti}},
  \bibinfo{author}{\bibfnamefont{B.}~\bibnamefont{Capogrosso-Sansone}},
  \bibnamefont{and}
  \bibinfo{author}{\bibfnamefont{M.}~\bibnamefont{Lewenstein}},
  \bibinfo{journal}{J. Phys. B: At, Mol. Opt. Phys.}
  \textbf{\bibinfo{volume}{44}}, \bibinfo{pages}{193001}
  (\bibinfo{year}{2011}).

\bibitem[{\citenamefont{Baranov et~al.}(2012)\citenamefont{Baranov, Dalmonte,
  Pupillo, and Zoller}}]{baranov}
\bibinfo{author}{\bibfnamefont{M.}~\bibnamefont{Baranov}},
  \bibinfo{author}{\bibfnamefont{M.}~\bibnamefont{Dalmonte}},
  \bibinfo{author}{\bibfnamefont{G.}~\bibnamefont{Pupillo}}, \bibnamefont{and}
  \bibinfo{author}{\bibfnamefont{P.}~\bibnamefont{Zoller}},
  \bibinfo{journal}{Chem. Rev.} \textbf{\bibinfo{volume}{112}},
  \bibinfo{pages}{5012} (\bibinfo{year}{2012}).

\bibitem[{\citenamefont{Lahaye et~al.}(2009)\citenamefont{Lahaye, Menotti,
  Santos, Lewenstein, and Pfau}}]{lahayephysics}
\bibinfo{author}{\bibfnamefont{T.}~\bibnamefont{Lahaye}},
  \bibinfo{author}{\bibfnamefont{C.}~\bibnamefont{Menotti}},
  \bibinfo{author}{\bibfnamefont{L.}~\bibnamefont{Santos}},
  \bibinfo{author}{\bibfnamefont{M.}~\bibnamefont{Lewenstein}},
  \bibnamefont{and} \bibinfo{author}{\bibfnamefont{T.}~\bibnamefont{Pfau}},
  \bibinfo{journal}{Rep. Prog. Phys.} \textbf{\bibinfo{volume}{72}},
  \bibinfo{pages}{126401} (\bibinfo{year}{2009}).

\bibitem[{\citenamefont{DeMille}(2002)}]{DeMille}
\bibinfo{author}{\bibfnamefont{D.}~\bibnamefont{DeMille}},
  \bibinfo{journal}{Phys. Rev. Lett.} \textbf{\bibinfo{volume}{88}},
  \bibinfo{pages}{067901} (\bibinfo{year}{2002}).

\bibitem[{\citenamefont{Zelevinsky et~al.}(2008)\citenamefont{Zelevinsky,
  Kotochigova, and Ye}}]{Zelevinsky}
\bibinfo{author}{\bibfnamefont{T.}~\bibnamefont{Zelevinsky}},
  \bibinfo{author}{\bibfnamefont{S.}~\bibnamefont{Kotochigova}},
  \bibnamefont{and} \bibinfo{author}{\bibfnamefont{J.}~\bibnamefont{Ye}},
  \bibinfo{journal}{Phys. Rev. Lett.} \textbf{\bibinfo{volume}{100}},
  \bibinfo{pages}{043201} (\bibinfo{year}{2008}).

\bibitem[{\citenamefont{Carr et~al.}(2009)\citenamefont{Carr, DeMille, Krems,
  and Ye}}]{carr:cold_2009}
\bibinfo{author}{\bibfnamefont{L.~D.} \bibnamefont{Carr}},
  \bibinfo{author}{\bibfnamefont{D.}~\bibnamefont{DeMille}},
  \bibinfo{author}{\bibfnamefont{R.~V.} \bibnamefont{Krems}}, \bibnamefont{and}
  \bibinfo{author}{\bibfnamefont{J.}~\bibnamefont{Ye}}, \bibinfo{journal}{New
  J. Phys.} \textbf{\bibinfo{volume}{11}}, \bibinfo{pages}{055049}
  (\bibinfo{year}{2009}).

\bibitem[{\citenamefont{Ospelkaus et~al.}(2010)\citenamefont{Ospelkaus, Ni,
  Wang, De~Miranda, Neyenhuis, Qu{\'e}m{\'e}ner, Julienne, Bohn, Jin, and
  Ye}}]{silke2010}
\bibinfo{author}{\bibfnamefont{S.}~\bibnamefont{Ospelkaus}},
  \bibinfo{author}{\bibfnamefont{K.-K.} \bibnamefont{Ni}},
  \bibinfo{author}{\bibfnamefont{D.}~\bibnamefont{Wang}},
  \bibinfo{author}{\bibfnamefont{M.}~\bibnamefont{De~Miranda}},
  \bibinfo{author}{\bibfnamefont{B.}~\bibnamefont{Neyenhuis}},
  \bibinfo{author}{\bibfnamefont{G.}~\bibnamefont{Qu{\'e}m{\'e}ner}},
  \bibinfo{author}{\bibfnamefont{P.}~\bibnamefont{Julienne}},
  \bibinfo{author}{\bibfnamefont{J.}~\bibnamefont{Bohn}},
  \bibinfo{author}{\bibfnamefont{D.~S.} \bibnamefont{Jin}}, \bibnamefont{and}
  \bibinfo{author}{\bibfnamefont{J.}~\bibnamefont{Ye}},
  \bibinfo{journal}{Science} \textbf{\bibinfo{volume}{327}},
  \bibinfo{pages}{853} (\bibinfo{year}{2010}).

\bibitem[{\citenamefont{Ni et~al.}(2010)\citenamefont{Ni, Ospelkaus, Wang,
  Qu{\'e}m{\'e}ner, Neyenhuis, De~Miranda, Bohn, Ye, and Jin}}]{ni2010}
\bibinfo{author}{\bibfnamefont{K.-K.} \bibnamefont{Ni}},
  \bibinfo{author}{\bibfnamefont{S.}~\bibnamefont{Ospelkaus}},
  \bibinfo{author}{\bibfnamefont{D.}~\bibnamefont{Wang}},
  \bibinfo{author}{\bibfnamefont{G.}~\bibnamefont{Qu{\'e}m{\'e}ner}},
  \bibinfo{author}{\bibfnamefont{B.}~\bibnamefont{Neyenhuis}},
  \bibinfo{author}{\bibfnamefont{M.}~\bibnamefont{De~Miranda}},
  \bibinfo{author}{\bibfnamefont{J.}~\bibnamefont{Bohn}},
  \bibinfo{author}{\bibfnamefont{J.}~\bibnamefont{Ye}}, \bibnamefont{and}
  \bibinfo{author}{\bibfnamefont{D.~S.} \bibnamefont{Jin}},
  \bibinfo{journal}{Nature} \textbf{\bibinfo{volume}{464}},
  \bibinfo{pages}{1324} (\bibinfo{year}{2010}).

\bibitem[{\citenamefont{De~Miranda et~al.}(2011)\citenamefont{De~Miranda,
  Chotia, Neyenhuis, Wang, Qu{\'e}m{\'e}ner, Ospelkaus, Bohn, Ye, and
  Jin}}]{miranda2011}
\bibinfo{author}{\bibfnamefont{M.}~\bibnamefont{De~Miranda}},
  \bibinfo{author}{\bibfnamefont{A.}~\bibnamefont{Chotia}},
  \bibinfo{author}{\bibfnamefont{B.}~\bibnamefont{Neyenhuis}},
  \bibinfo{author}{\bibfnamefont{D.}~\bibnamefont{Wang}},
  \bibinfo{author}{\bibfnamefont{G.}~\bibnamefont{Qu{\'e}m{\'e}ner}},
  \bibinfo{author}{\bibfnamefont{S.}~\bibnamefont{Ospelkaus}},
  \bibinfo{author}{\bibfnamefont{J.}~\bibnamefont{Bohn}},
  \bibinfo{author}{\bibfnamefont{J.}~\bibnamefont{Ye}}, \bibnamefont{and}
  \bibinfo{author}{\bibfnamefont{D.~S.} \bibnamefont{Jin}},
  \bibinfo{journal}{Nature Phys.} \textbf{\bibinfo{volume}{7}},
  \bibinfo{pages}{502} (\bibinfo{year}{2011}).

\bibitem[{\citenamefont{Yan et~al.}(2013)\citenamefont{Yan, Moses, Gadway,
  Covey, Hazzard, Rey, Jin, and Ye}}]{Bo2013}
\bibinfo{author}{\bibfnamefont{B.}~\bibnamefont{Yan}},
  \bibinfo{author}{\bibfnamefont{S.~A.} \bibnamefont{Moses}},
  \bibinfo{author}{\bibfnamefont{B.}~\bibnamefont{Gadway}},
  \bibinfo{author}{\bibfnamefont{J.~P.} \bibnamefont{Covey}},
  \bibinfo{author}{\bibfnamefont{K.~R.~A.} \bibnamefont{Hazzard}},
  \bibinfo{author}{\bibfnamefont{A.~M.} \bibnamefont{Rey}},
  \bibinfo{author}{\bibfnamefont{D.~S.} \bibnamefont{Jin}}, \bibnamefont{and}
  \bibinfo{author}{\bibfnamefont{J.}~\bibnamefont{Ye}},
  \bibinfo{journal}{Nature} \textbf{\bibinfo{volume}{501}},
  \bibinfo{pages}{521} (\bibinfo{year}{2013}).

\bibitem[{\citenamefont{Syassen et~al.}(2008)\citenamefont{Syassen, Bauer,
  Lettner, Volz, Dietze, Garc{\'\i}a-Ripoll, Cirac, Rempe, and
  D{\"u}rr}}]{rempe2008}
\bibinfo{author}{\bibfnamefont{N.}~\bibnamefont{Syassen}},
  \bibinfo{author}{\bibfnamefont{D.}~\bibnamefont{Bauer}},
  \bibinfo{author}{\bibfnamefont{M.}~\bibnamefont{Lettner}},
  \bibinfo{author}{\bibfnamefont{T.}~\bibnamefont{Volz}},
  \bibinfo{author}{\bibfnamefont{D.}~\bibnamefont{Dietze}},
  \bibinfo{author}{\bibfnamefont{J.}~\bibnamefont{Garc{\'\i}a-Ripoll}},
  \bibinfo{author}{\bibfnamefont{J.}~\bibnamefont{Cirac}},
  \bibinfo{author}{\bibfnamefont{G.}~\bibnamefont{Rempe}}, \bibnamefont{and}
  \bibinfo{author}{\bibfnamefont{S.}~\bibnamefont{D{\"u}rr}},
  \bibinfo{journal}{Science} \textbf{\bibinfo{volume}{320}},
  \bibinfo{pages}{1329} (\bibinfo{year}{2008}).

\bibitem[{\citenamefont{Mark et~al.}(2011)\citenamefont{Mark, Haller, Lauber,
  Danzl, Daley, and N\"agerl}}]{mark2011}
\bibinfo{author}{\bibfnamefont{M.~J.} \bibnamefont{Mark}},
  \bibinfo{author}{\bibfnamefont{E.}~\bibnamefont{Haller}},
  \bibinfo{author}{\bibfnamefont{K.}~\bibnamefont{Lauber}},
  \bibinfo{author}{\bibfnamefont{J.~G.} \bibnamefont{Danzl}},
  \bibinfo{author}{\bibfnamefont{A.~J.} \bibnamefont{Daley}}, \bibnamefont{and}
  \bibinfo{author}{\bibfnamefont{H.-C.} \bibnamefont{N\"agerl}},
  \bibinfo{journal}{Phys. Rev. Lett.} \textbf{\bibinfo{volume}{107}},
  \bibinfo{pages}{175301} (\bibinfo{year}{2011}).

\bibitem[{\citenamefont{Misra and Sudarshan}(1977)}]{misra:zeno_1988}
\bibinfo{author}{\bibfnamefont{B.}~\bibnamefont{Misra}} \bibnamefont{and}
  \bibinfo{author}{\bibfnamefont{E.~C.~G.} \bibnamefont{Sudarshan}},
  \bibinfo{journal}{J. Math. Phys.} \textbf{\bibinfo{volume}{18}},
  \bibinfo{pages}{756} (\bibinfo{year}{1977}).

\bibitem[{\citenamefont{Itano et~al.}(1990)\citenamefont{Itano, Heinzen,
  Bollinger, and Wineland}}]{itano:quantum_1990}
\bibinfo{author}{\bibfnamefont{W.~M.} \bibnamefont{Itano}},
  \bibinfo{author}{\bibfnamefont{D.~J.} \bibnamefont{Heinzen}},
  \bibinfo{author}{\bibfnamefont{J.~J.} \bibnamefont{Bollinger}},
  \bibnamefont{and} \bibinfo{author}{\bibfnamefont{D.~J.}
  \bibnamefont{Wineland}}, \bibinfo{journal}{Phys. Rev. A}
  \textbf{\bibinfo{volume}{41}}, \bibinfo{pages}{2295} (\bibinfo{year}{1990}).

\bibitem[{\citenamefont{Fischer et~al.}(2001)\citenamefont{Fischer,
  Guti\'errez-Medina, and Raizen}}]{PhysRevLett.87.040402}
\bibinfo{author}{\bibfnamefont{M.~C.} \bibnamefont{Fischer}},
  \bibinfo{author}{\bibfnamefont{B.}~\bibnamefont{Guti\'errez-Medina}},
  \bibnamefont{and} \bibinfo{author}{\bibfnamefont{M.~G.}
  \bibnamefont{Raizen}}, \bibinfo{journal}{Phys. Rev. Lett.}
  \textbf{\bibinfo{volume}{87}}, \bibinfo{pages}{040402}
  (\bibinfo{year}{2001}).

\bibitem[{\citenamefont{Han et~al.}(2009)\citenamefont{Han, Chan, Yi, Daley,
  Diehl, Zoller, and Duan}}]{Han2009}
\bibinfo{author}{\bibfnamefont{Y.-J.} \bibnamefont{Han}},
  \bibinfo{author}{\bibfnamefont{Y.-H.} \bibnamefont{Chan}},
  \bibinfo{author}{\bibfnamefont{W.}~\bibnamefont{Yi}},
  \bibinfo{author}{\bibfnamefont{A.~J.} \bibnamefont{Daley}},
  \bibinfo{author}{\bibfnamefont{S.}~\bibnamefont{Diehl}},
  \bibinfo{author}{\bibfnamefont{P.}~\bibnamefont{Zoller}}, \bibnamefont{and}
  \bibinfo{author}{\bibfnamefont{L.-M.} \bibnamefont{Duan}},
  \bibinfo{journal}{Phys. Rev. Lett.} \textbf{\bibinfo{volume}{103}},
  \bibinfo{pages}{070404} (\bibinfo{year}{2009}).

\bibitem[{\citenamefont{Garc{\'\i}a-Ripoll
  et~al.}(2009)\citenamefont{Garc{\'\i}a-Ripoll, D{\"u}rr, Syassen, Bauer,
  Lettner, Rempe, and Cirac}}]{Ripoll2009}
\bibinfo{author}{\bibfnamefont{J.~J.} \bibnamefont{Garc{\'\i}a-Ripoll}},
  \bibinfo{author}{\bibfnamefont{S.}~\bibnamefont{D{\"u}rr}},
  \bibinfo{author}{\bibfnamefont{N.}~\bibnamefont{Syassen}},
  \bibinfo{author}{\bibfnamefont{D.}~\bibnamefont{Bauer}},
  \bibinfo{author}{\bibfnamefont{M.}~\bibnamefont{Lettner}},
  \bibinfo{author}{\bibfnamefont{G.}~\bibnamefont{Rempe}}, \bibnamefont{and}
  \bibinfo{author}{\bibfnamefont{J.}~\bibnamefont{Cirac}},
  \bibinfo{journal}{New J. Phys.} \textbf{\bibinfo{volume}{11}},
  \bibinfo{pages}{013053} (\bibinfo{year}{2009}).

\bibitem[{\citenamefont{Baur and Mueller}(2010). (We note that in this
  reference there is a factor of 2 missing on the right hand side of Eq.
  (13))}]{baur:two-body_2010}
\bibinfo{author}{\bibfnamefont{S.~K.} \bibnamefont{Baur}} \bibnamefont{and}
  \bibinfo{author}{\bibfnamefont{E.~J.} \bibnamefont{Mueller}},
  \bibinfo{journal}{Phys. Rev. A} \textbf{\bibinfo{volume}{82}},
  \bibinfo{pages}{023626} (\bibinfo{year}{2010). (We note that in this
  reference there is a factor of 2 missing on the right hand side of Eq.
  (13)}).

\bibitem[{\citenamefont{Hazzard et~al.}(2013)\citenamefont{Hazzard, Manmana,
  Foss-Feig, and Rey}}]{Hazzard2013}
\bibinfo{author}{\bibfnamefont{K.~R.~A.} \bibnamefont{Hazzard}},
  \bibinfo{author}{\bibfnamefont{S.~R.} \bibnamefont{Manmana}},
  \bibinfo{author}{\bibfnamefont{M.}~\bibnamefont{Foss-Feig}},
  \bibnamefont{and} \bibinfo{author}{\bibfnamefont{A.~M.} \bibnamefont{Rey}},
  \bibinfo{journal}{Phys. Rev. Lett.} \textbf{\bibinfo{volume}{110}},
  \bibinfo{pages}{075301} (\bibinfo{year}{2013}).

\bibitem[{\citenamefont{Vidal}(2004)}]{Vidal2004}
\bibinfo{author}{\bibfnamefont{G.}~\bibnamefont{Vidal}},
  \bibinfo{journal}{Phys. Rev. Lett.} \textbf{\bibinfo{volume}{93}},
  \bibinfo{pages}{040502} (\bibinfo{year}{2004}).

\bibitem[{\citenamefont{Daley et~al.}(2004)\citenamefont{Daley, Kollath,
  Schollw\"ock, and Vidal}}]{Daley2004}
\bibinfo{author}{\bibfnamefont{A.~J.} \bibnamefont{Daley}},
  \bibinfo{author}{\bibfnamefont{C.}~\bibnamefont{Kollath}},
  \bibinfo{author}{\bibfnamefont{U.}~\bibnamefont{Schollw\"ock}},
  \bibnamefont{and} \bibinfo{author}{\bibfnamefont{G.}~\bibnamefont{Vidal}},
  \bibinfo{journal}{J. Stat. Mech. Theor. Exp.} p. \bibinfo{pages}{P04005}
  (\bibinfo{year}{2004}).

\bibitem[{\citenamefont{White and Feiguin}(2004)}]{White2004}
\bibinfo{author}{\bibfnamefont{S.~R.} \bibnamefont{White}} \bibnamefont{and}
  \bibinfo{author}{\bibfnamefont{A.~E.} \bibnamefont{Feiguin}},
  \bibinfo{journal}{Phys. Rev. Lett.} \textbf{\bibinfo{volume}{93}},
  \bibinfo{pages}{076401} (\bibinfo{year}{2004}).

\bibitem[{\citenamefont{Carmichael}(1991)}]{carmichael_open_1991}
\bibinfo{author}{\bibfnamefont{H.}~\bibnamefont{Carmichael}},
  \emph{\bibinfo{title}{An Open Systems Approach to Quantum Optics, Lectures
  Presented at the Universit\'e Libre de Bruxelles}}, Lecture Notes in Physics
  monographs (\bibinfo{publisher}{Springer}, \bibinfo{year}{1991}).

\bibitem[{\citenamefont{M{\o}lmer et~al.}(1993)\citenamefont{M{\o}lmer, Castin,
  and Dalibard}}]{molmer_monte_1993}
\bibinfo{author}{\bibfnamefont{K.}~\bibnamefont{M{\o}lmer}},
  \bibinfo{author}{\bibfnamefont{Y.}~\bibnamefont{Castin}}, \bibnamefont{and}
  \bibinfo{author}{\bibfnamefont{J.}~\bibnamefont{Dalibard}},
  \bibinfo{journal}{J. Opt. Soc. Am. B} \textbf{\bibinfo{volume}{10}},
  \bibinfo{pages}{524} (\bibinfo{year}{1993}).

\bibitem[{\citenamefont{Dum et~al.}(1992)\citenamefont{Dum, Parkins, Zoller,
  and Gardiner}}]{dum_monte_1992}
\bibinfo{author}{\bibfnamefont{R.}~\bibnamefont{Dum}},
  \bibinfo{author}{\bibfnamefont{A.~S.} \bibnamefont{Parkins}},
  \bibinfo{author}{\bibfnamefont{P.}~\bibnamefont{Zoller}}, \bibnamefont{and}
  \bibinfo{author}{\bibfnamefont{C.~W.} \bibnamefont{Gardiner}},
  \bibinfo{journal}{Phys. Rev. A} \textbf{\bibinfo{volume}{46}},
  \bibinfo{pages}{4382} (\bibinfo{year}{1992}).

\bibitem[{\citenamefont{Neyenhuis et~al.}(2012)\citenamefont{Neyenhuis, Yan,
  Moses, Covey, Chotia, Petrov, Kotochigova, Ye, and Jin}}]{brian}
\bibinfo{author}{\bibfnamefont{B.}~\bibnamefont{Neyenhuis}},
  \bibinfo{author}{\bibfnamefont{B.}~\bibnamefont{Yan}},
  \bibinfo{author}{\bibfnamefont{S.~A.} \bibnamefont{Moses}},
  \bibinfo{author}{\bibfnamefont{J.~P.} \bibnamefont{Covey}},
  \bibinfo{author}{\bibfnamefont{A.}~\bibnamefont{Chotia}},
  \bibinfo{author}{\bibfnamefont{A.}~\bibnamefont{Petrov}},
  \bibinfo{author}{\bibfnamefont{S.}~\bibnamefont{Kotochigova}},
  \bibinfo{author}{\bibfnamefont{J.}~\bibnamefont{Ye}}, \bibnamefont{and}
  \bibinfo{author}{\bibfnamefont{D.~S.} \bibnamefont{Jin}},
  \bibinfo{journal}{Phys. Rev. Lett.} \textbf{\bibinfo{volume}{109}},
  \bibinfo{pages}{230403} (\bibinfo{year}{2012}).

\bibitem[{fno({\natexlab{a}})}]{fnote1}
\emph{\bibinfo{title}{\textup{Note: Although a transition dipole matrix element
  exists between $\ket{\uparrow}$ and $\ket{\downarrow}$, we neglect dipolar
  interactions because of the lack of spin coherence in our 50:50 mixture}}}.

\bibitem[{\citenamefont{Chotia et~al.}(2012)\citenamefont{Chotia, Neyenhuis,
  Moses, Yan, Covey, Foss-Feig, Rey, Jin, and Ye}}]{chotia:long-lived_2012}
\bibinfo{author}{\bibfnamefont{A.}~\bibnamefont{Chotia}},
  \bibinfo{author}{\bibfnamefont{B.}~\bibnamefont{Neyenhuis}},
  \bibinfo{author}{\bibfnamefont{S.~A.} \bibnamefont{Moses}},
  \bibinfo{author}{\bibfnamefont{B.}~\bibnamefont{Yan}},
  \bibinfo{author}{\bibfnamefont{J.~P.} \bibnamefont{Covey}},
  \bibinfo{author}{\bibfnamefont{M.}~\bibnamefont{Foss-Feig}},
  \bibinfo{author}{\bibfnamefont{A.~M.} \bibnamefont{Rey}},
  \bibinfo{author}{\bibfnamefont{D.~S.} \bibnamefont{Jin}}, \bibnamefont{and}
  \bibinfo{author}{\bibfnamefont{J.}~\bibnamefont{Ye}}, \bibinfo{journal}{Phys.
  Rev. Lett.} \textbf{\bibinfo{volume}{108}}, \bibinfo{pages}{080405}
  (\bibinfo{year}{2012}).

\bibitem[{\citenamefont{Huang and Yang}(1957)}]{huang1957}
\bibinfo{author}{\bibfnamefont{K.}~\bibnamefont{Huang}} \bibnamefont{and}
  \bibinfo{author}{\bibfnamefont{C.~N.} \bibnamefont{Yang}},
  \bibinfo{journal}{Phys. Rev.} \textbf{\bibinfo{volume}{105}},
  \bibinfo{pages}{767} (\bibinfo{year}{1957}).

\bibitem[{fno({\natexlab{b}})}]{fnote}
\emph{\bibinfo{title}{\textup{Note: The sums run over all lattice sites,
  operators acting outside the bounds are implicitly ignored.}}}

\bibitem[{\citenamefont{Tsunetsugu et~al.}(1997)\citenamefont{Tsunetsugu,
  Sigrist, and Ueda}}]{Sigrist1997}
\bibinfo{author}{\bibfnamefont{H.}~\bibnamefont{Tsunetsugu}},
  \bibinfo{author}{\bibfnamefont{M.}~\bibnamefont{Sigrist}}, \bibnamefont{and}
  \bibinfo{author}{\bibfnamefont{K.}~\bibnamefont{Ueda}},
  \bibinfo{journal}{Rev. Mod. Phys.} \textbf{\bibinfo{volume}{69}},
  \bibinfo{pages}{809} (\bibinfo{year}{1997}).

\bibitem[{\citenamefont{Schachenmayer et~al.}(2013)\citenamefont{Schachenmayer,
  Pollet, Troyer, and Daley}}]{s_sp_2013}
\bibinfo{author}{\bibfnamefont{J.}~\bibnamefont{Schachenmayer}},
  \bibinfo{author}{\bibfnamefont{L.}~\bibnamefont{Pollet}},
  \bibinfo{author}{\bibfnamefont{M.}~\bibnamefont{Troyer}}, \bibnamefont{and}
  \bibinfo{author}{\bibfnamefont{A.~J.} \bibnamefont{Daley}},
  \bibinfo{journal}{arXiv: 1305.1301}  (\bibinfo{year}{2013}).

\bibitem[{\citenamefont{Foss-Feig et~al.}(2012)\citenamefont{Foss-Feig, Daley,
  Thompson, and Rey}}]{FossFeig2012}
\bibinfo{author}{\bibfnamefont{M.}~\bibnamefont{Foss-Feig}},
  \bibinfo{author}{\bibfnamefont{A.~J.} \bibnamefont{Daley}},
  \bibinfo{author}{\bibfnamefont{J.~K.} \bibnamefont{Thompson}},
  \bibnamefont{and} \bibinfo{author}{\bibfnamefont{A.~M.} \bibnamefont{Rey}},
  \bibinfo{journal}{Phys. Rev. Lett.} \textbf{\bibinfo{volume}{109}},
  \bibinfo{pages}{230501} (\bibinfo{year}{2012}).

\bibitem[{\citenamefont{Gorshkov et~al.}(2010)\citenamefont{Gorshkov, Hermele,
  Gurarie, Xu, Julienne, Ye, Zoller, Demler, Lukin, and Rey}}]{Gorshkov2010}
\bibinfo{author}{\bibfnamefont{A.}~\bibnamefont{Gorshkov}},
  \bibinfo{author}{\bibfnamefont{M.}~\bibnamefont{Hermele}},
  \bibinfo{author}{\bibfnamefont{V.}~\bibnamefont{Gurarie}},
  \bibinfo{author}{\bibfnamefont{C.}~\bibnamefont{Xu}},
  \bibinfo{author}{\bibfnamefont{P.}~\bibnamefont{Julienne}},
  \bibinfo{author}{\bibfnamefont{J.}~\bibnamefont{Ye}},
  \bibinfo{author}{\bibfnamefont{P.}~\bibnamefont{Zoller}},
  \bibinfo{author}{\bibfnamefont{E.}~\bibnamefont{Demler}},
  \bibinfo{author}{\bibfnamefont{M.}~\bibnamefont{Lukin}}, \bibnamefont{and}
  \bibinfo{author}{\bibfnamefont{A.}~\bibnamefont{Rey}},
  \bibinfo{journal}{Nature Phys.} \textbf{\bibinfo{volume}{6}},
  \bibinfo{pages}{289} (\bibinfo{year}{2010}).

\bibitem[{\citenamefont{Kuns et~al.}(2011)\citenamefont{Kuns, Rey, and
  Gorshkov}}]{Kuns2011}
\bibinfo{author}{\bibfnamefont{K.~A.} \bibnamefont{Kuns}},
  \bibinfo{author}{\bibfnamefont{A.~M.} \bibnamefont{Rey}}, \bibnamefont{and}
  \bibinfo{author}{\bibfnamefont{A.~V.} \bibnamefont{Gorshkov}},
  \bibinfo{journal}{Phys. Rev. A} \textbf{\bibinfo{volume}{84}},
  \bibinfo{pages}{063639} (\bibinfo{year}{2011}).

\bibitem[{\citenamefont{Baranov et~al.}(2005)\citenamefont{Baranov, Osterloh,
  and Lewenstein}}]{Lewenstein}
\bibinfo{author}{\bibfnamefont{M.~A.} \bibnamefont{Baranov}},
  \bibinfo{author}{\bibfnamefont{K.}~\bibnamefont{Osterloh}}, \bibnamefont{and}
  \bibinfo{author}{\bibfnamefont{M.}~\bibnamefont{Lewenstein}},
  \bibinfo{journal}{Phys. Rev. Lett.} \textbf{\bibinfo{volume}{94}},
  \bibinfo{pages}{070404} (\bibinfo{year}{2005}).

\bibitem[{\citenamefont{Ludlow et~al.}(2011)\citenamefont{Ludlow, Lemke,
  Sherman, Oates, Qu\'em\'ener, von Stecher, and Rey}}]{Ludlow2011}
\bibinfo{author}{\bibfnamefont{A.~D.} \bibnamefont{Ludlow}},
  \bibinfo{author}{\bibfnamefont{N.~D.} \bibnamefont{Lemke}},
  \bibinfo{author}{\bibfnamefont{J.~A.} \bibnamefont{Sherman}},
  \bibinfo{author}{\bibfnamefont{C.~W.} \bibnamefont{Oates}},
  \bibinfo{author}{\bibfnamefont{G.}~\bibnamefont{Qu\'em\'ener}},
  \bibinfo{author}{\bibfnamefont{J.}~\bibnamefont{von Stecher}},
  \bibnamefont{and} \bibinfo{author}{\bibfnamefont{A.~M.} \bibnamefont{Rey}},
  \bibinfo{journal}{Phys. Rev. A} \textbf{\bibinfo{volume}{84}},
  \bibinfo{pages}{052724} (\bibinfo{year}{2011}).

\bibitem[{\citenamefont{Bishof et~al.}(2011)\citenamefont{Bishof, Martin,
  Swallows, Benko, Lin, Qu\'em\'ener, Rey, and Ye}}]{Bishof2011b}
\bibinfo{author}{\bibfnamefont{M.}~\bibnamefont{Bishof}},
  \bibinfo{author}{\bibfnamefont{M.~J.} \bibnamefont{Martin}},
  \bibinfo{author}{\bibfnamefont{M.~D.} \bibnamefont{Swallows}},
  \bibinfo{author}{\bibfnamefont{C.}~\bibnamefont{Benko}},
  \bibinfo{author}{\bibfnamefont{Y.}~\bibnamefont{Lin}},
  \bibinfo{author}{\bibfnamefont{G.}~\bibnamefont{Qu\'em\'ener}},
  \bibinfo{author}{\bibfnamefont{A.~M.} \bibnamefont{Rey}}, \bibnamefont{and}
  \bibinfo{author}{\bibfnamefont{J.}~\bibnamefont{Ye}}, \bibinfo{journal}{Phys.
  Rev. A} \textbf{\bibinfo{volume}{84}}, \bibinfo{pages}{052716}
  (\bibinfo{year}{2011}).

\bibitem[{\citenamefont{Traverso et~al.}(2009)\citenamefont{Traverso,
  Chakraborty, Martinez~de Escobar, Mickelson, Nagel, Yan, and
  Killian}}]{Traverso2009}
\bibinfo{author}{\bibfnamefont{A.}~\bibnamefont{Traverso}},
  \bibinfo{author}{\bibfnamefont{R.}~\bibnamefont{Chakraborty}},
  \bibinfo{author}{\bibfnamefont{Y.~N.} \bibnamefont{Martinez~de Escobar}},
  \bibinfo{author}{\bibfnamefont{P.~G.} \bibnamefont{Mickelson}},
  \bibinfo{author}{\bibfnamefont{S.~B.} \bibnamefont{Nagel}},
  \bibinfo{author}{\bibfnamefont{M.}~\bibnamefont{Yan}}, \bibnamefont{and}
  \bibinfo{author}{\bibfnamefont{T.~C.} \bibnamefont{Killian}},
  \bibinfo{journal}{Phys. Rev. A} \textbf{\bibinfo{volume}{79}},
  \bibinfo{pages}{060702} (\bibinfo{year}{2009}).

\bibitem[{\citenamefont{Lisdat et~al.}(2009)\citenamefont{Lisdat, Winfred,
  Middelmann, Riehle, and Sterr}}]{Lisdat2009}
\bibinfo{author}{\bibfnamefont{C.}~\bibnamefont{Lisdat}},
  \bibinfo{author}{\bibfnamefont{J.~S. R.~V.} \bibnamefont{Winfred}},
  \bibinfo{author}{\bibfnamefont{T.}~\bibnamefont{Middelmann}},
  \bibinfo{author}{\bibfnamefont{F.}~\bibnamefont{Riehle}}, \bibnamefont{and}
  \bibinfo{author}{\bibfnamefont{U.}~\bibnamefont{Sterr}},
  \bibinfo{journal}{Phys. Rev. Lett.} \textbf{\bibinfo{volume}{103}},
  \bibinfo{pages}{090801} (\bibinfo{year}{2009}).

\bibitem[{\citenamefont{\.{Z}uchowski and Hutson}(2010)}]{PhysRevA.81.060703}
\bibinfo{author}{\bibfnamefont{P.~S.} \bibnamefont{\.{Z}uchowski}}
  \bibnamefont{and} \bibinfo{author}{\bibfnamefont{J.~M.}
  \bibnamefont{Hutson}}, \bibinfo{journal}{Phys. Rev. A}
  \textbf{\bibinfo{volume}{81}}, \bibinfo{pages}{060703}
  (\bibinfo{year}{2010}).

\bibitem[{\citenamefont{Busch et~al.}(1998)\citenamefont{Busch, Englert,
  Rza{\.z}ewski, and Wilkens}}]{Butsch2000}
\bibinfo{author}{\bibfnamefont{T.}~\bibnamefont{Busch}},
  \bibinfo{author}{\bibfnamefont{B.-G.} \bibnamefont{Englert}},
  \bibinfo{author}{\bibfnamefont{K.}~\bibnamefont{Rza{\.z}ewski}},
  \bibnamefont{and} \bibinfo{author}{\bibfnamefont{M.}~\bibnamefont{Wilkens}},
  \bibinfo{journal}{Found. Phys.} \textbf{\bibinfo{volume}{28}},
  \bibinfo{pages}{549} (\bibinfo{year}{1998}).

\bibitem[{\citenamefont{Verstraete et~al.}(2008)\citenamefont{Verstraete, Murg,
  and Cirac}}]{verstraete2008}
\bibinfo{author}{\bibfnamefont{F.}~\bibnamefont{Verstraete}},
  \bibinfo{author}{\bibfnamefont{V.}~\bibnamefont{Murg}}, \bibnamefont{and}
  \bibinfo{author}{\bibfnamefont{J.~I.} \bibnamefont{Cirac}},
  \bibinfo{journal}{Advances in Physics} \textbf{\bibinfo{volume}{57}},
  \bibinfo{pages}{143 } (\bibinfo{year}{2008}).

\bibitem[{\citenamefont{Schollw\"ock}(2011)}]{schollwock2011}
\bibinfo{author}{\bibfnamefont{U.}~\bibnamefont{Schollw\"ock}},
  \bibinfo{journal}{Annals of Physics} \textbf{\bibinfo{volume}{326}},
  \bibinfo{pages}{96} (\bibinfo{year}{2011}), ISSN \bibinfo{issn}{0003-4916}.

\end{thebibliography}

\begin{widetext}
\appendix*{SUPPLEMENTARY MATERIAL}

\section{I A. Renormalization of the on-site loss rate by higher bands}  As stated in Eq.~(2) of the main text, the single-band approximation predicts an on-site loss rate $\Gamma_0$ that is proportional to the loss rate coefficient $\beta^{(3D)}$.  When $\hbar \beta^{3D} a^3$, $a$ the lattice spacing, becomes of the order of the band gap, this approximation breaks down due to renormalization of the Wannier orbitals by higher band mixing, and $\Gamma_0$ is likewise renormalized.  The renormalization of $\Gamma_0$ is shown in Fig.~S\ref{fig:SupplSingleWell}(a), which shows the numerically computed on-site loss rate for two particles in a single well of an optical lattice with depth $V=50E_R$.  Here, the chemical reactions are parameterized in terms of an inelastic $s$-wave scattering length $a_s$ as $\hbar \beta^{3D}=4\pi \hbar^2 a_s/m$, and the units are the effective harmonic oscillator frequency $\hbar \omega=2\sqrt{V E_R}$ and length $a_{\mathrm{ho}}=(a/\pi)(V/E_R)^{-1/4}$ of the lattice well.  The single-band result (red) continues to increase as losses increase, but the multiband result (blue) initially increases, reaches a maximum, and then begins to decrease.  Some physical insight into this non-monotonic behavior can be obtained by comparing to two particles in a harmonic oscillator with elastic contact interactions, where analytic analysis is possible~\cite{Butsch2000}.  The analog of the loss rate for elastic interactions is the interaction energy, which peaks when $a_s/a_{\mathrm{ho}}\sim 1$.  Then, as elastic interactions become very strong, $a_s/a_{\mathrm{ho}}\to\infty$, the probability for two particles to be at the same position goes to zero due to mixing in of higher bands.  Because interactions occur only when two particles come in contact, the diminished probability for the two particles to be close means that the interaction energy decreases, just as our loss rate decreases.  The dashed vertical line in Fig.~S\ref{fig:SupplSingleWell}(a) indicates the parameters for KRb which is deep in the regime where the Wannier orbitals are strongly renormalized.

\section{I B. Renormalization of the double-well effective loss rate by higher bands} In the main text it was shown that the effective loss rate for two particles in a double well potential with tunneling $J$ and on-site loss rate $\Gamma_0$ is $4\Gamma_{\mathrm{eff}}$, $\Gamma_{\mathrm{eff}}=2\left(J/\hbar\right)^2/\Gamma_0$, within second-order perturbation theory and the single-band approximation.  In the previous section, we showed that the on-site loss rate $\Gamma_0$ is strongly renormalized by the inclusions of higher bands.  In addition, $J$ is strongly renormalized when higher-band effects are included.  The numerically obtained effective loss rate as a function of the transverse lattice confinement is displayed in Fig. S\ref{fig:SupplSingleWell}(b) using the loss rate coefficient $\beta_{\mathrm{3D}}$ for KRb and a fixed lattice depth $V_y=5E_R$ for the tunneling direction.  The number of bands used in the calculation increases from top to bottom.  As $V_{\perp}$ is increased, the loss rate within the lowest band is increased due to stronger localization of the Wannier functions.  The fact that the single-band calculation, the uppermost line in in Fig. S\ref{fig:SupplSingleWell}(b), predicts an increasing effective loss rate with increasing $V_{\perp}$ hence clearly demonstrates that $J$ is strongly renormalized by losses.  When higher bands are included, we see a marked qualitative change in which the effective loss rate decreases as $V_{\perp}$ is increased, in accordance with the single-band Zeno expectation.  In addition, we see that the number of bands required for convergence is roughly the same for all transverse lattice heights.  This is due to the fact that the interactions that renormalize the single-particle Wannier functions and the band gap have identical scaling with $V_{\perp}$ in the deep-lattice limit.

\begin{suppfigure*}[!htbp]
  \begin{center}
 \includegraphics[width=120mm]{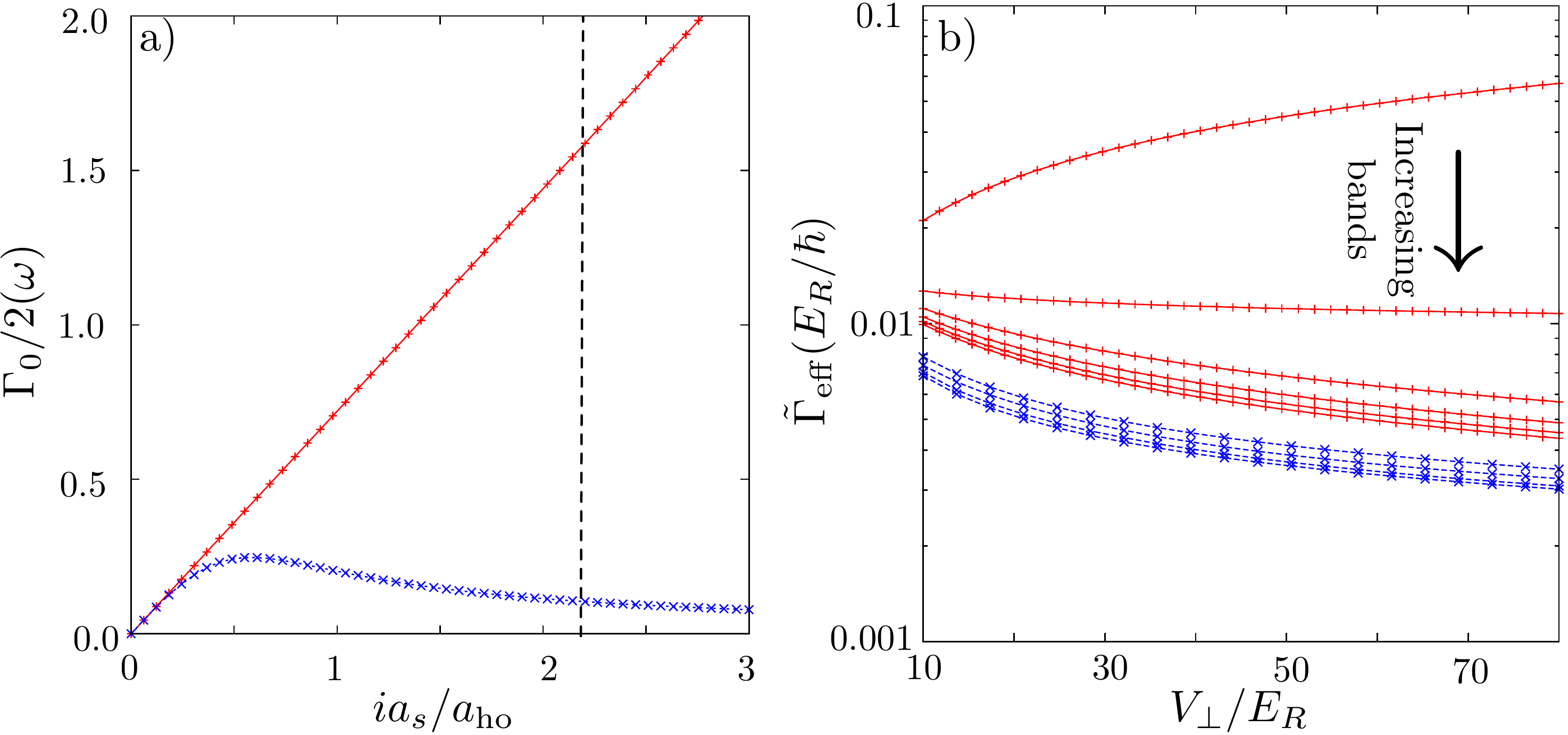}
 \end{center}
 \caption{(Color online) 
(a) The on-site loss rate in oscillator units ($\hbar \omega=2\sqrt{V E_R}$ is approximately the band gap) for two atoms in a single lattice site with depth $V=50E_R$ is shown for a single-band calculation (solid red line) and a calculation with $6^3$ bands (dashed blue line) as a function of the pure imaginary $s$-wave scattering length $a_s$ in oscillator units ($a_{\mathrm{ho}}=(a/\pi)(V/E_R)^{-1/4}$).  The vertical dotted line indicates the parameters for KRb.  (b) The effective loss rate of the double well calculation versus the transverse lattice height and the number of bands used in the calculation.  The red solid curves use a single transverse band and 1 to 6 bands along the tunneling direction from top to bottom.  The blue dashed curves use 6 bands along the tunneling direction and $3^2$ to $6^2$ bands along the transverse directions from top to bottom.
   }
\label{fig:SupplSingleWell}
\end{suppfigure*}
\section{II. numerical models for dissipative dynamics}
We model the dissipative dynamics  by means of a master equation for the density operator $\hat{\rho}$:
\begin{align}
\label{master}  \frac{d}{dt} {\hat \rho} &= -\frac{{\rm i}}{\hbar}[ {\hat H_0} , {\hat \rho} ] +  {\mathcal L}{\hat\rho},
\end{align}
with $\hat{H}_0$ describing the unitary evolution and  ${\mathcal L}{\hat \rho}=\frac{1}{2}\sum_j(\hat{\rho}\hat{C}_j^\dagger\hat{C}_j+\hat{C}_j^\dagger\hat{C}_j\hat{\rho}-2\hat{C}_j\hat{\rho}\hat{C}_j^\dagger)$ accounting for dissipation. In a 1D tube with a superimposed  lattice potential, $\hat{H}_0=-J\sum_{j,\sigma}  (\hat{c}^\dagger_{j\sigma }\hat{c}_{j+1\sigma} + h.c.)+\hat{V}_{\text{trap}}$, where $J$ represents the hopping of molecules between sites, and $\hat{V}_{\text{trap}}$ represents the parabolic trapping potential. The two-body losses that occur between KRb molecules are described by jump operators $\hat{C}_j=\sqrt{ \Gamma_0 }\hat{c}_{j\uparrow}\hat{c}_{j\downarrow}$, which destroy a pair of molecules in different rotational states ($\ket{\uparrow}$ or $\ket{\downarrow}$) at the same site $j$, and $\Gamma_0$ is the on-site loss rate. In the regime $\Gamma_0\gg J/\hbar$, the doubly occupied states can be adiabatically eliminated~\cite{Ripoll2009}, projecting the Hilbert space to states with at most one molecules per site. As a result, the jump operators $\hat{C}_j$ are replaced by $ \hat{A}_j=\sqrt{2\Gamma_{\text{eff}}}\Big[(\hat{c}_{j\uparrow} \hat{c}_{j+1\downarrow}+\hat{c}_{j\uparrow}\hat{c}_{j-1\downarrow})-(\hat{c}_{j\downarrow}\hat{c}_{j+1\uparrow}+\hat{c}_{j\downarrow}\hat{c}_{j-1\uparrow})\Big]$, and $\Gamma_{\text{eff}}=2(J/\hbar)^2/\Gamma_0$. These new jump operators $\hat{A}_j$ effectively remove spin singlet states formed in neighoring sites, and account for second order tunneling processes. Even within the reduced space, the above master equation is generally difficult to solve for experimentally relevant lattice sizes. In 1D, however, various numerical techniques can be applied. In the following we provide the details of the numerical methods utilized in this paper for solving the above master equation.  
 
\subsection{A. t-DMRG}
To numerically exactly simulate the dynamics of Eq.~\eqref{master} in 1D, we make use of a {\em matrix product state} (MPS) ansatz \cite{verstraete2008,schollwock2011}. In contrast to the {\em product state} factorization that is used in the mean-field treatment (see below), this ansatz allows for a finite entanglement entropy between different bi-partitions of the state. The amount of entropy that can be captured depends on the MPS bond-dimension, $\chi_{\rm MPS}$. In this way, for large enough $\chi_{\rm MPS}$ it becomes possible to exactly represent pure quantum states that are not greatly entangled. The application of operators, such as they appear as jump operators in the dissipative part of  Eq.~\eqref{master}  to a MPS can be readily implemented \cite{schollwock2011}. In addition, well-established t-DMRG methods \cite{Vidal2004,Daley2004,White2004} can be used to simulate (non-hermitian) Hamiltonian dynamics of a MPS. With this ability together with the possibility to rewrite the master equation evolution in terms of random samplings of such trajectories~\cite{carmichael_open_1991,molmer_monte_1993,dum_monte_1992}, it becomes possible to calculate statistical estimates for the time-evolution of simple observables, such as the number of $\ket{\downarrow}$ molecules. Besides the statistical convergence for the Monte Carlo sampling, we repeated simulations with increasingly large $\chi_{\rm MPS}$ to obtain convergence in the bond-dimension, i.e.~to ensure that the full amount of entanglement that is produced in the dynamics is captured.
\subsection{B. Mean-field treatment (MF)}
The above master equation can be significantly simplified by adopting a MF ansatz that allows for efficient numerical treatment of realistic system sizes. Under this ansatz, $\hat{\rho}=\prod_j \tilde{\rho}_{j}$ with $\tilde{\rho}_{j} \equiv \sum_{\alpha,\beta=\{\uparrow,\downarrow,0\}} {\rho}_j^{\alpha,\beta} |\alpha\rangle \langle \beta|$. $\uparrow,\downarrow,0$ label the three possible local states of spin up, spin down, and the vacuum, respectively. Solving the master equation  reduces to solving a set of coupled and non-linear differential equations for $\rho_j^{\alpha,\beta}$:
\begin{align}
\frac{d\rho_j^{\sigma\sigma}}{dt}&=\frac{i}{\hbar}J\sum_{l}
(\rho_l^{\sigma0}\rho_j^{0\sigma}-\rho_l^{0\sigma}\rho_j^{\sigma0})+2\Gamma_{\text{eff}}\sum_{l}(\rho_l^{\sigma'\sigma}\rho_j^{\sigma\sigma'}+\rho_l^{\sigma\sigma'}\rho_j^{\sigma'\sigma})-4\Gamma_{\text{eff}}\sum_{l}\rho_l^{\sigma'\sigma'}\rho_j^{\sigma\sigma},\label{eq:m1}
\\
   \frac{d\rho_j^{\sigma\sigma'}}{dt}&= \frac{i}{\hbar}(\Omega_\sigma-\Omega_{\sigma'})j^2\rho_j^{\sigma\sigma'}+\frac{i}{\hbar}J\sum_{l}
 (\rho_l^{\sigma0}\rho_j^{0\sigma'}-\rho_l^{0\sigma'}\rho_j^{\sigma0})+2\Gamma_{\text{eff}}\sum_{l}\!\sum_{\alpha}\!(\rho_l^{\sigma\sigma'}\rho_j^{\alpha\alpha}-\rho_l^{\alpha\alpha}\rho_j^{\sigma\sigma'}),
 \\
    \frac{d\rho_j^{0\sigma}}{dt}&=\frac{i}{\hbar}\Omega_\sigma j^2\rho_j^{0\sigma}+\frac{i}{\hbar}J\sum_{l}(\sum_{\alpha}\rho_l^{0\alpha}\rho_j^{\alpha\sigma}-\rho_l^{0\sigma}\rho_j^{00})+2\Gamma_{\text{eff}}\sum_{l}(\rho_l^{\sigma'\sigma}\rho_j^{0\sigma'}-\rho_l^{\sigma'\sigma'}\rho_j^{0\sigma}),
 \\
    \frac{d\rho_j^{\sigma0}}{dt}&=-\frac{i}{\hbar}\Omega_\sigma j^2\rho_j^{\sigma0}+\frac{i}{\hbar}J\sum_{l}(\rho_l^{\sigma0}\rho_j^{00}-\sum_{\alpha}\rho_l^{\alpha0}\rho_j^{\sigma\alpha})+2\Gamma_{\text{eff}}\sum_{l}(\rho_l^{\sigma\sigma'}\rho_j^{\sigma' 0}-\rho_l^{\sigma'\sigma'}\rho_j^{\sigma0}),
 \\
  \frac{d\rho_j^{00}}{dt}&= \frac{i}{\hbar}J\sum_l\sum_{\alpha}(\rho_l^{0\alpha}\rho_j^{\alpha0}-\rho_l^{\alpha0}\rho_j^{0\alpha})+4\Gamma_{\text{eff}}\sum_{\mathclap{l}}\!\sum_{\alpha\not=\alpha'}\!\sum_{\beta\not=\beta'}\!(-1)^{\rm{\delta}_{\alpha\beta}}\rho_l^{\alpha\beta}\rho_j^{\alpha'\beta'},\label{eq:m5}
\end{align}
where $\sigma, \alpha, \alpha',
\beta, \beta'\in\{\uparrow,\downarrow\}$, $\sigma\neq\sigma'$, the summation of $l$ is over the nearest neighbors of $j$, and $\Omega_{\uparrow(\downarrow)}=\frac{1}{2}m\omega_{\uparrow(\downarrow)}^2a^2$ with $a$ the lattice spacing. In deriving the above Eqs.~\eqref{eq:m1}-\eqref{eq:m5}, we have neglected terms such as $\hat{n}_{j\sigma} \hat{c}^\dagger_{j+1\sigma '}\hat{c}_{j-1\sigma '}\hat{\rho}$ and $\hat{c}^\dagger_{j+1\sigma}\hat{c}_{j\sigma}\hat{c}^\dagger_{j\sigma '}\hat{c}_{j-1\sigma '}\hat{\rho}$ in ${\mathcal L}\hat{\rho}$, which correspond to the correlated hopping processes involving three sites. We have performed calculations to confirm the contributions from these terms are small for the systems we treat in this paper.

\subsection{\label{sec:supIIC}C. Rate equation (RE)}
A further simplification can be obtained by neglecting the dynamical redistribution of molecules and the effect of inhomogeneity, as well as coherences. This is equivalent to dropping the unitary evolution terms in Eqs.~\eqref{eq:m1}-\eqref{eq:m5}, and setting all off-diagonal terms of the density matrix $\tilde{\rho}_j$ to 0. With these approximations, for a 50:50 mixture of molecules, $\rho_j^{\uparrow\uparrow}=n_{\uparrow}=\rho_{j+1}^{\downarrow\downarrow}=n_{\downarrow}$,  Eq.~\eqref{eq:m1}  simplifies to    
\begin{align}
\frac{dn_{\downarrow}}{dt}=-8\Gamma_{\text{eff}} [n_{\downarrow}(t)]^2,
\end{align}
which is the same as the RE Eq.~(3) 
  in the main text. The RE gives a fairly good understanding of the loss dynamics at short times, as discussed in the main text. However, a quantitative estimate of the filling fraction requires going beyond these approximations. This can be accomplished by using the MF treatment.

\section{III. Long time saturation of molecule losses}
In our experiment the molecule number per tube is finite and dissipation stops in tubes where all the remaining molecules are in the same rotational state. As a result, the loss saturates  with time. Signatures of this  effect are observed at long times in the experiment, as shown in Fig.~S\ref{fig:figlong}. The simplified RE cannot capture the saturation, but it does describe the experimental data at short times. We extract the initial loss rate by fitting to times during which $\sim$25\% molecules are lost, and the RE  fits well [Fig.~S\ref{fig:figlong}], especially for the relatively shallow lattice depths
 \begin{suppfigure*}
  \begin{center}
 \includegraphics[width=70mm]{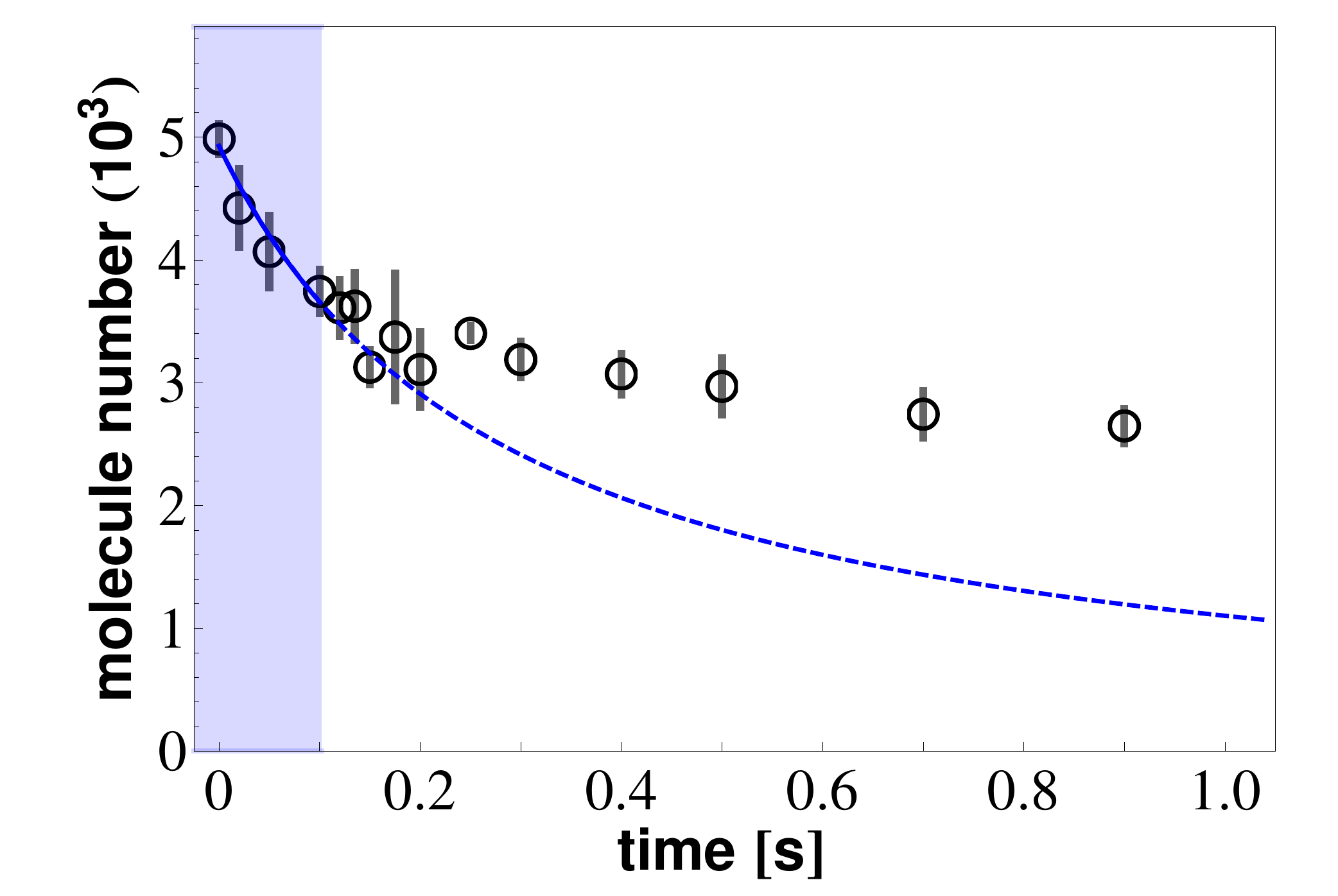}
 \end{center}
 \caption{(Color online) 
Molecule loss vs time for lattice depth of $V_y=4.6~E_R$ and $V_\perp=40~E_R$: the experimental data (black circles) exhibit saturation at long time, as a result of the finite molecule number per tube. The shaded area shows the time window used to fit the RE. During it only $\sim 25\%$ of the molecules have been lost. The RE fit (blue solid line)  breaks down at longer times (blue dashed line).
}
\label{fig:figlong}
\end{suppfigure*}
\end{widetext}

\end{document}